\documentclass{aa}
\usepackage{graphics}
\begin{document}
\thesaurus
	{07 (03.20.2; 	      
	     03.20.8;	      
	     08.09.2 P~Cyg;   
	     08.13.2)         
	}
\title{The relative figure of merit of optical interferometry as compared to spectroscopy }
\subtitle{Example of parameter estimates for the circumstellar 
envelope of P~Cyg}
\author{M.S.~Burgin\inst{1,2}
        \and 
        A.~Chalabaev\inst{2}}
\offprints{A.~Chalabaev}
\institute{Astro-Space Center FIAN, Profsoyuznaya 84/32, 117810 Moscow, 
Russia, mburgin@dpc.asc.rssi.ru 
           \and
           CNRS, Laboratoire d'Astrophysique, Observatoire de 
           Grenoble, UMR 5571, BP 53X, F-38041 Grenoble 
           CEDEX, France, Almas.Chalabaev@obs.ujf-grenoble.fr
           }
\date{Received date: June 14, 1999
}
\titlerunning{Interferometry vs Spectroscopy, Case of P~Cyg wind}
\authorrunning{M.S.Burgin \& A. Chalabaev}
\maketitle
\begin{abstract}
When planning new facilities, one is interested to know whether and by 
how much the new technique is superior to already existing ones.  We 
describe a general approach permitting to evaluate the relative merits of various
techniques used in astrophysical observations, following the theory of 
model parameter estimation.  It is applied to compare 
two-aperture optical long baseline interferometry to classical 
spectroscopy, both used to define the model parameters of the
P~Cyg wind.  The wind modeling was done using an efficient 
approximation for computation of the line source function; it allowed 
to analyse about $10^5$ points in the parametric space of P~Cyg 
envelope models.  It is shown that interferometry offers no real 
advantage if the object can be described by stationary spherically 
symmetric models with a priori fixed thermal structure.  However, if 
the object must be described by a model with a large number of free 
parameters, e.g. when the thermal structure of the envelope is not 
fixed a priori, then the interferometric measurements can reduce the 
error in model parameters determination by an order of magnitude.  In 
the case of P~Cyg, the reduction of errors provided by interferometry 
is highest for the baseline lengths in the range 45--90 m.  This 
illustrates the capacity of the proposed method to be used for 
optimization of interferometric configurations.  The  
simplifications adopted for this first trial application are 
critically analyzed, and future improvements are indicated.
\end{abstract}
\section{Introduction}
When evaluating justification for new facilities, one may question 
whether and by how much the new technique is superior to already 
existing ones.  We describe a general approach which leads to a 
quantitative answer, expressed as a relative figure of merit of two 
observational techniques.  The approach is based on the comparison of 
the corresponding errors in the model parameters determination.  The 
derived figure of merit can also be used in the closely related 
problem of optimization of complex observations, thus contributing in 
maximizing their scientific output.

We illustrate the developed method by comparing the technique of 
optical long baseline interferometry to the classical spectroscopy, 
both used to find the parameters of the gaseous envelope around the P 
Cyg star.  For this first trial application, we use simulated data, 
obtained by solving the radiative transfer problem for grids of 
physical models of the envelope and then adding the photon noise.

The chosen example of the 
interferometry vs spectroscopy comparison appears to be timely.  
Indeed, in the recent years, interferometry has been
undergoing a steady progress witnessed by publication of fringe 
visibilities for circumstellar envelopes of various types sometimes 
combined with a good spectral resolution (Mourard et al.  
\cite{mour89}, Vakili et al.  \cite{iau162vakili}, \cite{gi2tpcyg}; 
Quirrenbach et al.  \cite{quir93}, \cite{quir94}, \cite{quir97}; 
Harmanec et al.  \cite{harmanec}; Stee et al.  \cite{stee}; see also 
expected performances of interferometric arrays under construction, 
VLTI, von der L\"uhe et al. \cite{luehe}, Petrov et al. \cite{amber}, 
and CHARA, McAlister et al.  \cite{chara}).  While this effort is 
fully justified, for the knowledge of the flux distribution on small angular scales 
can unveil important new features of astrophysical objects, it is 
also justified to wonder whether this effort is well payed for all 
classes of objects and for all kinds of physical problems.  As a matter of fact, 
interferometric observations as compared to those at a single 
telescope are more complex, more expensive, more time consuming, and, 
when possible to compare, have a larger accumulated measurement error 
due to their complexity.  Furthermore, in the foreseeable future the 
instruments of this type will remain orders of magnitude less numerous
and by all given reasons much less available for the community
than single telescopes. Thus, it makes sense to evaluate
in quantitative terms what are the targets for which the
interferometric instrumentation has the highest potential in terms of
astrophysical information as compared to single telescopes.

Also, the optimization of interferometric observations may be a 
crucial aspect.  Indeed, the existing and forthcoming long baseline 
interferometers comprise only few individual apertures, covering a 
small set of spatial frequencies during one observing cycle.  On the 
other hand, the information provided by observations can depend 
critically on spatial frequencies.  Certainly, in such a situation the 
scientific yield of the interferometric array could be greatly 
improved if thorough model calculations can indicate the 
configurations expected to be the most useful for determination of 
aimed physical parameters of the object.

Not surprisingly, the relative figure of merit derived in the present 
article depends not only on the experimental techniques, but also on 
the chosen set of model parameters, reflecting the fact that the 
physical model makes a necessary part of the merit problem. 
However, in modern astronomy, designing 
and building instrumentation on the one hand, and modeling objects on 
the other hand, were pushed to such a level of sophistication that 
they are as a rule two distinct activities.  The optimum use of 
interferometry needs a close collaboration of two communities.

After these preliminary remarks, let us specify that in the present 
first trial application, our discussion will be limited to the 
comparison of observations with a two-aperture optical long baseline 
interferometer (hereafter OLBI) and the classical single telescope 
spectroscopy (hereafter shortly spectroscopy), both applied to study a 
circumstellar gaseous envelope formed by the wind of a massive star.  
The OLBI observations are assumed to have the same spectral resolution 
as spectroscopy.
The discussion will be further limited to spectral profiles and 
visibilities in the H$\alpha$ line of the P~Cyg envelope, 
allowing to use published high quality observations both in OLBI and 
spectroscopic, and thus to provide a clear numerical illustration.

In our earlier attempts to solve the OLBI vs spectroscopy merits 
problem (Burgin \& Chalabaev \cite{notre1}; Bourguine \& Chalabaev 
\cite{notre2}), we used qualitative comparison of observables, 
visibility and spectral profile, computed for a small number of 
envelope models.  The results were ambiguous.  Analising them, we 
arrived to the firm conviction that meaningful conclusions can be 
obtained only if (1) intercomparison of different types of 
observations is performed using a clearly defined figure of merit, 
indicating how much information on the studied object is provided by 
various techniques, and (2) a sufficiently large ranges in the space of 
envelope model parameters are analyzed.  

In the case of P~Cyg, the 
required number of computed models (see Sect.~\ref{results}) turned out to 
be $10^4-10^5$.  An efficient simplified envelope model code,
computing the emergent spectral profile and visibilities for the 
H$\alpha$ line, was developed and is described in 
Sect.~\ref{model}.  We considered only spherically symmetric outflow 
models, the choice which is justified in Sect.~\ref{model}.

\section{Relative figure of merit of observational methods \label{statth}}
\subsection{Definitions and general theory. \label{general-theory}}
We consider astrophysical observations and their interpretation in the 
framework of the theory of parameter estimation (e.g. Kendall \& Stuart \cite{ks}).  That is, the 
observed object is assumed to be exactly described by a fixed 
multiparametric physical model, the errors being solely those of the 
experimental measurements (see however Sect.~\ref{systerrs}).  Any 
observable quantity is then a function of the vector of {$M_{\mathrm 
P}$} model parameters
$\vec \Theta=(\Theta_1,\Theta_2,\ldots,\Theta_{M_{\mathrm P}})$.

An observation consists in finding the values of $M_{\mathrm O}$ 
observables 
$\vec{\hat Y}=(\hat Y_1,\hat Y_2,\ldots,\hat Y_{M_{\mathrm O}})$,
where each observable $\hat Y_i$ is either measured directly or can be 
calculated in a model-independent way as a known function of directly 
measurable quantities.  Obviously, observations with instruments of 
different types provide observables of different number and nature.

The $i$-th observable can be represented as
       $\hat Y_i=Y_i(\vec\Theta)+\varepsilon_i$,
where
       $Y_i(\vec\Theta)$ is the ``theoretical'' value
of corresponding observable in the absence of observational errors and
    $\vec\varepsilon
     =(\varepsilon_1,\varepsilon_2,\ldots,\varepsilon_{M_{\mathrm 
     O}})$ 
is the random vector of observational errors, which is assumed here to 
be distributed normally.  We consider the case when $M_{\mathrm 
O}>M_{\mathrm P}$ and the parameters $\vec\Theta$ could have been 
uniquely determined if the vector $\vec Y$ were known.

We shall call the parameters to be determined the ``target parameters'' 
and denote a set of {$M_{\mathrm T}$} target parameters, or target 
set, as $\mathcal T$.  It is possible that $M_{\mathrm T}<M_{\mathrm 
P}$ and the set {$\mathcal T$} is a proper subset of the set of all 
parameters of the model.  The situation of this kind may arise for two 
reasons.
 
First, a model parameter may be set to some a priori fixed value 
because, say, it has been measured earlier by entirely different 
methods.  This reduces the dimension of the problem and computational 
difficulties, the general method remaining the same.  Let {$\mathcal 
F$} denote the set of {$M_{\mathrm F}$} fixed parameters, and 
$\tilde\Theta_f$ be the value a priori assigned to a parameter 
$\Theta_f\in\mathcal F$.

Secondly, some of the unknown physical parameters may eventually be 
considered ``not interesting'' if their values are not relevant to the 
astrophysical problem under study.  Following the theory of parameter 
estimation, we will call them the ``nuisance'' 
parameters and denote the set of nuisance parameters, or ``nuisance set'',
as {$\mathcal N$}, their 
number being $M_{\mathrm N}$.  As a rule, their influence can be 
separated in the error analysis only at the final stage of 
computations, for their values are to be calculated along with the 
values of target parameters.

Interpretation of an observation consists in calculating the vector 
${\vec{\hat \Theta}}_{\mathrm T}(\vec{\hat Y}) = (\hat 
\Theta_{t(1)},\hat \Theta_{t(2)},\ldots, \hat \Theta_{t(M_{\mathrm 
T})})$
that provides the best fit to the observed values \vec{\hat Y}.
Here $t(i)$ is the index of the component of vector $\Theta$
corresponding to the same model parameter as the $i$-th target
parameter ($i=1,\ldots,M_{\mathrm T}$).

The precision of the values $\vec{\hat \Theta}_{\mathrm T}$ resulting from
interpretation of the observations is characterized by 
  ${\tens M}(\vec{\hat \Theta}_{\mathrm T}-\vec\Theta_{\mathrm T})$
and
  ${\tens D}\vec{\hat \Theta}_{\mathrm T}$,
where for any random vector  $\vec\xi$ expressions 
${\tens M}\vec\xi$ and ${\tens D}\vec\xi$ 
denote its mean and its covariance matrix respectively,
and $\vec\Theta_{\mathrm T}=
     (\Theta_{t(1)}, \Theta_{t(2)},\ldots, \Theta_{t(M_{\mathrm T})})$

In what follows, we assume that $\vec{\hat Y}$ is an unbiased estimate 
of $\vec Y$, i.e.  ${\tens M}{\vec\varepsilon}=0$, and that the 
measurement errors are small enough, so that the error analysis can be 
performed using the linearized version of the least squares method.

We additionally assume that all a priori fixed parameters are set to 
their true values (see however \ref{systerrs}), that is 
$\tilde\Theta_f=\Theta_f$ for any $\Theta_f\in\mathcal F$, then
${\tens M}{\vec{\hat 
\Theta}=\vec\Theta}$
and statistical properties of errors in parameter determination are 
completely characterized by the covariance matrix of errors
\tens C = {\tens D}\vec{\hat \Theta}.
\subsection{Random errors and the relative figure of merit\label{randerr}}
According to Kendall \& Stuart (\cite[Chap.19]{ks}), the covariance 
matrix of errors \tens C is related to $\vec Y$ and $\vec \Theta$ by the 
following expression:
\begin{equation}
\tens C(\vec\Theta,\mathcal T,\mathcal N)
     =\left({\tens A}^{\mathrm T} 
            {\left({\tens D}\vec\varepsilon\/\right)}^{-1} 
      \tens A \right)^{-1}
                                \label{covmateval}      
\end{equation}
where 
$\tens A$ 
is the 
$M_{\mathrm O}\times (M_{\mathrm T}+M_{\mathrm N})$ 
matrix with elements 
$ A_{ij} = \frac  {\partial Y_i} {\partial \Theta_j} $
for $\Theta_j\in\mathcal T\cup\mathcal N$ and $\vec\varepsilon$ is the
vector of experimental errors defined in Sect.~\ref{general-theory}.

The natural scalar characteristics of the precision of parameter 
determination for a given set {$\mathcal T$} of target parameters is 
the following principal subdeterminant of the covariance matrix:
\begin{eqnarray}                                        \label{Cdef}
\lefteqn{C(\vec\Theta,\mathcal T,\mathcal N)=}&&\\
&=&\det
\left|
\begin{array}{llcl}
 \tens C_{t(1)t(1)}&\tens C_{t(1)t(2)}&\ldots&\tens C_{t(1)t(M_{\mathrm T})}\\
 \tens C_{t(2)t(1)}&\tens C_{t(2)t(2)}&\ldots&\tens C_{t(2)t(M_{\mathrm T})}\\
 \vdots            & \vdots           &\ddots& \vdots                       \\ 
 \tens C_{t(M_{\mathrm T})t(1)}&\tens C_{t()t(2)}&\ldots&
                                  \tens C_{t(M_{\mathrm T})t(M_{\mathrm T})}
\end{array}
\right| \nonumber
\end{eqnarray}

Geometrically, this subdeterminant is proportional to the hypervolume 
of the scattering ellipsoid in the space of target parameters.  It 
depends not only on the physical model and errors of measurements, but 
also on the analytical form used for description of the model: two
physically equivalent but mathematically different
(e.g. interrelated by a reversible substitution of variables)
analytical representations could yield entirely different values
of $C(\vec\Theta,\mathcal T,\mathcal N)$.

However, a pair of observational methods can well be compared if one 
makes use of the ratio
\begin{equation}                                \label{Rdef}
R(\vec\Theta,\mathcal T,\mathcal N)=
 \left(
         C^{{\rm I}}(\vec\Theta,\mathcal T,\mathcal N) / 
          C^{{\rm II}}(\vec\Theta,\mathcal T,\mathcal N)
 \right)^{1/M_{\mathrm T}},
\end{equation}
where $C^{{\rm I}}$ and $C^{{\rm II}}$ are the subdeterminants 
calculated for the observation of the same object with instruments 
{{\rm I}} and {{\rm II}}.  This ratio depends only on the physical 
model used, and the obtained precision on the model parameters.  We 
shall call the quantity $R(\vec\Theta,\mathcal T,\mathcal N)$, 
introduced by the Eq.~(\ref{Rdef}), the ``random error ratio'', or 
the ``relative figure of merit''. 

If the instruments are of the same kind, and differ from each other 
only in precision, the value $R(\vec\Theta,\mathcal T,\mathcal N)$ is 
merely the ratio of observational errors.  If the instruments are of 
different kinds, providing observables of different nature and number, 
and the target set consists of only one model parameter, the value 
$R(\vec\Theta,\mathcal T,\mathcal N)$ is merely the ratio of resulting 
random errors in the parameter determination.  However, in the general 
case of instruments of arbitrary kinds and multiparametric models, no 
simple ratio of errors exists, and the evaluation of relative merits 
can be done only using the quantity $R(\vec\Theta,\mathcal T,\mathcal 
N)$ defined by Eq.~(\ref{Rdef}).

It allows to compare various 
observational techniques applied to objects described by various 
multiparametric models: equality 
$R(\vec\Theta,\mathcal T,\mathcal N)<1$ means that the instrument I is
better suited for determination of parameters from the target set than the
instrument II.
\subsection{Systematic errors induced by interpretation\label{systerrs}}
The fact that the description of an object by a multiparametric 
physical model is only an approximation to the reality implies that we 
have to consider the robustness of the method, that is the
stability of the results it yields with respect to deviations of
the real situation from the model.

The present framework offers a way to obtain certain quantitative
characteristics of the robustness.  
Indeed, let us consider a multiparametric model with $\mathcal 
F\neq\emptyset$.  If $\tilde\Theta_f\neq\Theta_f$ for some 
$\Theta_f\in\mathcal F$ then, in general, ${\tens 
M}{\hat\Theta_t}-\Theta_t \neq 0$ for $\Theta_t\in\mathcal T$.  That 
is, in addition to random errors of observational origin, the result 
is biased by systematic errors due to inaccurate
interpretation. The value of that bias characterizes the robustness of the
method with respect to deviations of $\tilde\Theta_f$ from its
true value.

In the linear approximation,
${\tens M}(\hat\Theta_t-\Theta_t)
 =\tens S\times(\tilde\Theta_f-\Theta_f)
$,
where 
$\tens S$ 
is the 
$M_{\mathrm T}\times M_{\mathrm F}$ 
matrix with elements
\begin{displaymath}                     
  \tens S_{tf}=
   \frac  {\partial \Theta_t} {\partial \Theta_f}\,,
\end{displaymath}
where $\Theta_t\in\mathcal T,\quad\Theta_f\in\mathcal F$.

A comprehensive study of systematic errors requires a joint analysis 
of individual elements of the matrix {$\tens S$} and uncertainties in 
model parameters from $\mathcal F$.  In the present paper, we will 
develop a simplified approach providing semiquantitative indications 
concerning the relative robustness of different observational 
techniques.

Let us first define for each observational method under consideration 
the value
\begin{equation}                                 \label{CtftRatio}
U(\Theta,\mathcal T,\mathcal F)
 =(C(\Theta,\mathcal T,\mathcal F)/C(\Theta,\mathcal T,\emptyset))^{1/M_{\mathrm T}}\,.
\end{equation}
When the estimates for target parameters are uncorrelated with 
estimates for parameters from {$\mathcal F$}, equality $U=1$ takes
place: the method is robust (with respect to deviations of
specified form from the model).  When such a correlation exists, $U$ 
exceeds unity; the closer the correlation, the larger its value, 
eventually implying poor robustness of the method.

Further, to compare the robustness of methods {{\rm I}} and {{\rm II}} 
with respect to inaccuracies in parameters from $\mathcal F$, we will 
define the ``robustness ratio $S$'' as follows:
\begin{equation}                                   \label{Sdef}
S(\vec\Theta,\mathcal T,\mathcal F)
 =U^{{\rm I}}/U^{{\rm II}}
 =R(\vec\Theta,\mathcal T,\mathcal F)/R(\vec\Theta,\mathcal T,\emptyset)
\end{equation}
where function $R$ is defined in Eq.~(\ref{Rdef}).  Note that once 
the values of $R(\Theta,\mathcal T,\mathcal N)$, which are necessary 
for analysis of random errors, are found, comparison of robustness is 
straightforward.
\section{The relative figure of merit of OLBI vs spectroscopy}
In this section we obtain the expressions for the elements of the 
covariance matrix ${\tens D}\vec{\hat \Theta}$ in the form that can be 
directly used to compare the OLBI and spectroscopy.

First, the measurement errors have to be specified.  For this first 
trial application, the only considered source of random errors is the 
photon noise.  Other sources of error, in particular those in OLBI 
arising from atmospheric seeing and calibrations of the modulation 
transfer function (see Roddier \& L\'ena \cite {roddier}; Mourard et 
al.  \cite {mour94}) are important and has to be incorporated in a 
future work.

Also, the spectral coverage and the spectral resolution have to be 
specified.  We assume that they are the same for both kinds of 
observations; the covered spectral region is $[\lambda_1,\lambda_2]$, 
the spectral resolution is sufficiently high, i.e. the monochromatic 
intensity received from each point of the observed object remains 
nearly constant across each spectral channel. 
In other words, we assume that $\Delta\lambda$, 
the width of the spectral channels, in 
terms of velocity is lesser than the thermal velocity of the emitting 
atoms.  For the P~Cyg type envelopes this corresponds to 
$\Delta\lambda\la 0.3$\,\AA{}, which is easily satisfied with modern 
spectrographs.  Then one can formally consider the convenient limit 
case $\Delta\lambda\to 0$.  In this case, the resulting covariance 
matrices tend to a limit which is independent of the spectral 
resolution (cf. Eqs.~(\ref{cspec}) and (\ref{colbi})).

\subsection{Spectroscopy \label{spectroscopy}}
The physical quantity provided by spectroscopy is the flux density as 
a function of wavelength.  In terms of observables, it is given by the 
vector ${\vec{\hat Y}}\!{}^{\mathrm S}$ with the components
\begin{equation}                                    \label{Ys}
 \hat Y^{\mathrm S}_i=\hat N_i,\ i=1,2,\ldots,L\,,
\end{equation}
where $\hat N_i$ is the number of photons recorded in the 
{\it i}-th spectral channel during the exposure, and $L$ is the number 
of channels.  The random values $\hat N_i$ are related to the
physical parameters of the object by
\begin{equation}                                  \label{Nest}
 \hat N_i=N_i+n_i\,,
\end{equation}
where
\begin{equation}                               \label{Nspec}
  N_i=F(\vec\Theta,\lambda_i) E \Delta_i \lambda,
\end{equation}
$F(\Theta,\lambda)$ is the monochromatic flux from the object at 
wavelength $\lambda$,  {$\Delta_i \lambda$} is the width of $i$-th 
spectral channel, and $n_i$ is the measurement error due to the photon 
noise, with the variance given by
\begin{equation}                               \label{Nerr}
  {\tens D}{n_i}=N_i\,.
\end{equation}
Finally the factor $E$ is given by
\begin{displaymath}
E=A_\mathrm{eff}t\,,
\end{displaymath}
where $A_\mathrm{eff}$ is the instrument effective aperture and
$t$ is the exposure time.
Hereafter we suppose that
$E^\mathrm{S} = E^\mathrm{I}$, where $E^\mathrm{S}$ and $E^\mathrm{I}$
pertain correspondingly to the spectrometer and interferometer.
Since the values $E^\mathrm{S}$ and $E^\mathrm{I}$ enter
the final results only through the ratio 
$E^\mathrm{S}/E^\mathrm{I}$,
we can for the sake of simplicity set 
$E^\mathrm{S} = E^\mathrm{I} = 1$.

Thus, 
  $\varepsilon^{\mathrm S}_i=n_i$
and the elements of the covariance matrix of measurement errors are
given by 
\begin{equation}
{\tens M}{(\varepsilon^{\mathrm S}_i \varepsilon^{\mathrm S}_j)}=
   \left\{
    \begin{array}{ll}
       0                & \mbox{if $ i \neq j $} \\
       N_i              & \mbox{if $ i=j $}               
    \end{array}
   \right.\,.                          \label{nncovar}
\end{equation}
Substituting Eqs.~(\ref{Ys})--(\ref{nncovar}) into 
Eq.~(\ref{covmateval})
and proceeding to the limit $\Delta_i \lambda\to 0$ we obtain:
\begin{equation}                            \label{cspec}         
  \left( \tens C^{\mathrm S}\right) ^{-1}_{pq} =
   \int_{\lambda_1}^{\lambda_2} \frac{1}{F(\vec\Theta,\lambda)}  
            \frac {\partial {F(\vec\Theta,\lambda)}} {\partial  \Theta_{p}} 
            \frac {\partial {F(\vec\Theta,\lambda)}} {\partial  \Theta_{q}}
          \,{\mathrm d}\lambda\,.                          
\end{equation}
For spherically symmetric objects considered in the present paper,
\begin{equation}                               \label{Fdef}
 F(\Theta,\lambda)
  =2\pi\int_0^\infty I(\Theta,\lambda,p)\,{\mathrm d}p\,,
\end{equation}
where  $I(\vec\Theta,\lambda,p)$ is the monochromatic intensity  at the
angular distance $p$ from the center of the object.
\subsection{OLBI \label{interferometry}}
We consider the case of a two-aperture OLBI with an adjustable baseline. The physical 
quantity eventually provided by OLBI measurements is the fringe 
visibility V as a function of the baseline vector.  The visibility V is equal 
to the real part of the complex degree of coherence, and therefore is 
the Fourier transform of the brightness distribution in the focal 
plane by virtue of the Van Cittert - Zernicke theorem (e.g. Mariotti 
\cite{mariotti}, textbooks of Pe\^rina \cite {perina}, and Goodman 
\cite {goodman}).  However, the value estimated in practice is 
$V^{2}$.  We refer the reader for details to the thorough discussion 
by Mourard et al.  (\cite {mour94}). In the present analysis we 
consider the estimates $\hat W_i$ of the quantity
\begin{equation}                         \label{Wdef}
W_i=W(\vec\Theta,\lambda_i,B)=V^2(\vec\Theta,\lambda_i,B), 
\end{equation}
where $i$ refers as previously to the $i$-th spectral channel, and $B$ 
is the projected baseline length of the interferometer.  
In the case of spherically symmetric objects
the intensity distribution is circularly symmetric and
$V$ is given by the 
normalized Hankel transform as follows (Bracewell \cite {bracewell}):
\begin{equation}                                     \label{Vaxisym}
 V(\vec\Theta,\lambda,B)
  =\frac
    {2\pi\int_0^{\infty} I(\vec\Theta,\lambda,p)J_0(kp)\,{\mathrm d}p}
    {F(\Theta,\lambda)}\,,
\end{equation}
where $J_0$ is the Bessel function, and $k=2 \pi B/\lambda$.

Statistical errors $w_i=\hat W_i - W_i$ obey the equations
\begin{equation}
{\tens M}{(w_i w_j)}=
   \left\{
    \begin{array}{ll}
       0                              & \mbox{if $ i \neq j $} \\
       {\tens D}(\hat W_i)= \frac{4W_i}{N_i}(2-W_i) 
                                      & \mbox{if $ i=j $}               
    \end{array}
   \right.,                          \label{wwcovar}
\end{equation}
and
\begin{equation}
{\tens M}{(w_i n_j)}=0\,.                   \label{nwcovar}
\end{equation}
The validity of Eqs.~(\ref{wwcovar}) and (\ref{nwcovar}) for $i \neq 
j$ is evident.  The case $i=j$ is treated in \ref{olbidiag}.

As a rule, the spectrum of the object, i.e. the values $\hat N_i$, is
also recorded.  Consequently, the vector of observables for the OLBI 
and the elements of its covariance matrix of errors are given by
\begin{equation}
\vec{\hat Y}^{\mathrm I}\label{Yolbi}
 =(\hat{N_1},\hat{W_1},\hat{N_2},\hat{W_2},\ldots,\hat{N_L},\hat{W_L})
\end{equation}
and
\begin{displaymath}
{\tens M}(\varepsilon^{\mathrm I}_i\varepsilon^{\mathrm I}_j)
  =\left\{
    \begin{array}{ll}
       0                         & \mbox { if $ i \neq j   $}  \\
       N_k                       & \mbox{ if $ i=j=2k-1   $}  \\
       \frac{4W_k}{N_k}(2-W_k)   & \mbox{ if $ i=j=2k     $} 
    \end{array}
   \right.\,.
\end{displaymath}
Substituting this expressions into Eq.~(\ref{covmateval}),
we obtain in the limit $\Delta_i \lambda\to 0$ that
\begin{equation}                                  \label{colbi}
  \left(\tens C^{\mathrm I}\right) ^{-1}_{pq}
  = \left(\tens C^{\mathrm S}\right) ^{-1}_{pq} +
    \int_{\lambda_1}^{\lambda_2}
             \frac{F}{4W(2-W)} \frac {\partial {W}} {\partial  \Theta_{p}} 
             \frac {\partial {W}} {\partial  \Theta_{q}}
           \,{\mathrm d}\lambda\,,
\end{equation}
where $\left(\tens C^{\mathrm S}\right) ^{-1}_{pq}$ is 
defined in Eq.~(\ref{cspec}).
\subsection{The relative figure of merit}        \label{RelFigMer}
As discussed in Sect.~\ref{randerr}, the relative figure of merit of 
OLBI as compared to spectroscopy is the random error ratio:
\begin{equation}
R^\mathrm{IS}(\vec\Theta,\mathcal T,\mathcal N,B)              \label{risdef}
 =C^{\mathrm I}(\vec\Theta,\mathcal T,\mathcal N,B)/C^{\mathrm S}(\vec\Theta,\mathcal T,\mathcal N)\,.
\end{equation}
From Eqs.~(\ref{colbi}) and (\ref{risdef}), it follows that 
$R^\mathrm{IS}<1$, which merely reflects the fact that the 
observables of spectroscopy constitute a subset of the OLBI 
observables.  Consequently, in the adopted comparison, the statistical 
errors in parameter determination by OLBI are always lesser than those 
obtained when only spectroscopic data are used, if however the 
spectroscopic data are of equal precision.

As explained in Sect. \ref{spectroscopy}, the Eq. (\ref{risdef})
is obtained in the assumption that 
$E^\mathrm{S} = E^\mathrm{I}$. To compare a pair of instruments
for which that equality does not hold, the RHS of
Eq. (\ref{risdef}) should be multiplied by 
$(E^\mathrm{S}/E^\mathrm{I})^{1/2}$.
\section{Models of the P~Cyg wind. Computed observables \label{model}}
\subsection{Stellar parameters \label{stpars}}
The star P~Cyg (B1 Iape) is characterized by a high mass-loss rate and 
exhibits bright optical emission lines, formed in the dense and nearly 
fully ionized stellar wind.  It belongs to the class of Luminous Blue 
Variables, a short-living transition phase in the evolution of a 
massive star at the end of the hydrogen burning (see e.g. {Humphreys} 
\& {Davidson} \cite{hd}, {Langer} et al.  \cite {langer}, {Maeder} 
\cite {maeder}).

There were at least two strong reasons to choose P~Cyg as the 
astrophysical object of the present study.  First, this bright object, 
$V = 4.8^{m}$, is one of the best studied emission line stars.  
Although in this first trial we use simulated data, it is important to 
note that for P~Cyg
there exists a rich literature providing not only high quality 
spectroscopic data (e.g. {Scuderi} et al.  \cite{scuderi}), but also 
interferometric data (Vakili et al.  \cite{gi2tpcyg}), as well as 
thorough theoretical analysis (Drew \cite{drew}; {Pauldrach} \& {Puls} 
\cite{pp}).  This insures that the present work can be followed up by 
a practical application.  Secondly, it happens that in the growing list 
of outflows known to be non-spherical (e.g. Wolf, Stahl, Fullerton 
\cite{iau169}), that of P~Cyg is an exception exhibiting spherical 
symmetry to a good degree of accuracy (Nota \cite{nota}), becoming 
clumpy only on short time and small flux scales ({Taylor} et al.  
\cite{asymwind}; Vakili et al.  \cite{gi2tpcyg}; Nota \cite{nota}).  
This implies that the assumption of spherical symmetry of the 
envelope, used in the present work, is realistic.  Let us remind that 
it allows to compute a model in a reasonable amount of time, and thus 
to explore a large domain of parametric space, which is a requirement 
for the evaluation results to be meaningful.

For the distance to the star $d$, its radius {$R_*$} and the effective 
temperature {$T_\mathrm{eff}$} we adopted the following values: 
$d=1800\,\mathrm{pc}$, $R_*=76.0\,R_\odot$, 
$T_\mathrm{eff}=20000\,{\mathrm K}$ (Lamers et al.  \cite{lgc}; 
Pauldrach \& Puls 
\cite{pp}).  The spectrum of the star was assumed to be blackbody.  
Our study is limited to the hydrogen H$\alpha$ line, 
which is the most prominent and the best studied feature in the 
spectrum of the star.
\subsection
 {The physical model of the envelope         \label{physmodl}} 
The mass distribution within a spherically symmetric and stationary 
outflow is described by its mass loss rate $\dot M$ and the outflow 
velocity $v$, which we assume to depend on radial distance $r$ in the 
following way:
\begin{displaymath}                       
v(r)=v_{\mathrm c}+(v_{\infty}-v_{\mathrm c})(1-1/r)^\alpha,
\end{displaymath}
where $v_{\infty}$ is the terminal velocity of the wind, $v_{\mathrm c}$ is
velocity of the wind at the base of the photosphere, and $\alpha$ is a
dimensionless parameter that characterizes the rate at which the velocity
approachs its asymptotic value at large $r$.
The radial distance $r$ is measured in units of stellar radius 
{$R_*$}, while $\dot M$, $v_{\mathrm c}$, $v_{\infty}$, and $\alpha$ 
are parameters of a model.

The temperature in the envelope is often assumed to be constant.  
However, as it was shown by Drew (\cite{drew}), across the region of 
the H$\alpha$ line formation $(R\loa5)$ the temperature 
in the envelope can vary by as much as 6000\,K. Therefore, along with 
isothermal models, we also computed the emergent emission for 
non-isothermal models.

To keep a finite number of scalar parameters, the dependence of 
envelope temperature $T(r)$ on radial distance is approximated in the 
following manner: it is assumed that $T(r_i)=T_i$ for 
$r_1=1<r_2<\dots<r_{M_R}$, that $T(r)$ is a linear function of $\log 
r$ at each interval $[r_i,r_{i+1}]$ for $0<i<{M_R}-1$, and that 
$T(r)=T_{M_R}$ for $r>r_{M_R}$.  The values $r_i$ were fixed for each 
model of the family.

We consider only non-increasing temperature laws $T(r)$, that is 
$\Delta_iT\geq0$ for $i=2,\dots,{M_R}$, where $\Delta_iT=T_{i-1}-T_i$.  
Since the computations are organized in such a way that the model 
parameters vary independently on each other, it is the values 
$\Delta_iT$ that are used along with $T_1$ as the parameters defining 
the envelope temperature, and the vector of model parameters is given 
by
\begin{displaymath}
\vec\Theta=(\alpha,\dot M,v_{\infty},v_{\mathrm c},T_1,\Delta_2T,
            \dots,\Delta_{{M_R}}T).
\end{displaymath}
\subsection{Source function and Radiative transfer equation}
The ``supersonic'' Sobolev approximation is a well known efficient 
method of calculating line spectral profiles (e.g. Castor 
\cite{castor}).  However, it would have a low accuracy if applied in 
the P~Cyg case, for a noticeable fraction of H$\alpha$ 
flux is emitted in the inner part of envelope, where the outflow 
velocity is comparable to that of sound.  We adopted a 
mixed method computing the source function in the Sobolev 
approximation, and then finding the emergent intensities by exact 
numerical solution of the transfer equation.  As it was shown
by Hamann (\cite{hamann}), the errors in line profiles computed in the 
original Sobolev approximation come mainly from calculations of the 
emergent intensities, whereas the source function is accurate for a 
wide range of physical conditions.

The choice of the model of the hydrogen atom was based on the fact 
that the regions of the envelope emitting the major fraction of the 
H$\alpha$ flux are nearly completely ionized and opaque 
to Lyman continuum and L$\alpha$, so that direct 
recombinations to and photoionizations from the ground level cancel 
out, and L$\alpha$ is saturated (Drew \cite{drew}).  The 
populations of levels $n=2$ and $n=3$ are mainly defined by 
collisional and radiative transitions between these two levels, 
radiative ionizations due to stellar radiation, and radiative 
recombinations (including indirect) to the level $n=3$.  We therefore 
adopted the three-level + continuum model of hydrogen atom.

The balance equations for $n_2$ and $n_3$,  the number densities of 
hydrogen atoms respectively at levels 2 and 3, take then the following 
form:
\begin{equation}                                            \label{balance}
\begin{array}{rlcrlcl}
n_2&(C_{23}+I_2)&-&n_3&(C_{32}+\beta_{32} A_{32})
   &=&R_2\,,\\
-n_2&C_{23}&+&n_3&(C_{32}+\beta_{32} A_{32}+I_3)
   &=&R_3\,,\\
\end{array}
\end{equation}
where $A_{32}$ is the spontaneous emission coefficient, $C_{23}$ and 
$C_{32}$ are collisional excitation and de-excitation coefficients for 
the transition, $I_2, I_3$ and $R_2, R_3$ are respectively ionizations 
coefficients and recombination rates for the levels involved, and 
$\beta_{32}$ is the escape probability.

Since the hydrogen is nearly completely ionized, the values $C_{23}$, 
$C_{32}$, $R_2$ and $R_3$ can be considered independent on $n_2$ and 
$n_3$, the only source of non-linearity in Eqs.~(\ref{balance}) being 
the terms containing $\beta_{23}$.

The escape probability $\beta_{32}$ is a rather complex function of 
the population of the lower level of the transition, so that commonly 
the Eqs.~(\ref{balance}) are solved by iterations, and this and, in 
particular, the recalculation of the escape probability, is by far the 
most time consuming operation.

To speed up computations, we developed a new approximate method for 
solving the equations of statistical equilibrium.  It is based on the 
fact that, as shown in \ref{epapprox}, the function $\beta_{32}(n_2)$ can 
be approximated with sufficiently good accuracy by a simple analytic 
expression as follows:
\begin{equation}
\beta_{32}(n_2)=1/(1+n_2/n_{\mathrm{as}}),          \label{betapp}
\end{equation}
where
\begin{equation}                           \label{nashalpha}
n_{\mathrm{as}}=  \frac{8\pi}{\lambda^3A_{32}} \frac{g_2}{g_3}
          \left(
            \frac{1}{3}\frac{{\mathrm d}v(r)}{{\mathrm d}r} +
            \frac{2}{3}\frac{v(r)}{r}
          \right),
\end{equation}
$g_2$ and $g_3$ are statistical weights of the levels, and $\lambda$
is the line wavelength. 

Introducing the dimensionless variables
\begin{equation}                       \label{Pdef}
P_2=n_2/n_0\,,\quad P_3=n_3/(n_0Q)\,,
\end{equation}  
where
\begin{displaymath}
n_0=\frac{R_2}{I_2},\quad 
Q=\frac{g_3}{g_2}\exp\left(-\frac{E_{23}}{kT}\right)\,,
\end{displaymath}
and $E_{23}$ is the energy of transition,
and substituting the approximation (\ref{betapp}) into 
Eqs.~(\ref{balance}) we obtain after some algebra that $P_2$ is
the positive solution of the quadric equation
\begin{displaymath}
aP_2^2+bP_2+c=0\,,
\end{displaymath}
where
\begin{eqnarray*}
a&=&-\rho_2MS\,,\\
b&=&(\rho_2*(1-\beta_0*(\delta+S))+M*(1-\beta_0)+\rho_3)P\,,\\
c&=&\beta_0*((\rho_2+\rho_3)*(1+\delta)+M)=0\,.
\end{eqnarray*}
Here
\begin{eqnarray*}
\rho_i&=&R_i/(A_{32}n_0)\quad{\rm for}\quad i=2,3\,,\\
M&=&\iota\rho_2^2\delta\,,\\
S&=&1+\iota Q\,,\\
\beta_0&=&1/(1+n_0/n_{\mathrm{as}})\,.
\end{eqnarray*}
and
\begin{displaymath}
\iota=I_3/I_2\,.
\end{displaymath}
When $P_2$ is calculated, the value of $P_3$ can be found using
the relation
\begin{displaymath}
P_3=\frac{QP_2+\rho_3\delta}{Q(\iota\rho_2\delta+\beta\delta+1)}\,.
\end{displaymath}
The source function is then easily computed from level
populations, which can be obtained using Eq.~(\ref{Pdef}).

Integration of the transfer equation was performed using the code 
developed by Bertout(\cite{bertout}), who kindly provided it to the 
authors.  In computing the line profile, the code assumes the envelope 
to be isothermal, so that its use for a non-isothermal case requires 
some comments.  The envelope temperature enters the calculations at 
two points: (1) In calculating the source function, through 
coefficients $C_{23}$ and $C_{32}$ of Eq.~(\ref{balance}).  Since the 
Bertout code is applicable to arbitrary source function, variations in 
temperature does not cause any difficulties here.  (2) In integrating 
the transfer equation, $T(r)$ enters the result through the local 
Doppler line width.  Since we consider only the envelopes with 
relatively low temperature contrast $(T_1-T_{{M_R}})/T_1\la0.4$, and 
dependence of the Doppler width on temperature is rather week, the 
error induced by this isothermal code structure is still negligible.
\subsection{Computation of final results \label{compfinres}}
Once the intensity $I(\vec\Theta,\lambda,p)$ is computed, the 
observable quantities, i.e. the emergent flux $F(\vec\Theta,\lambda)$ 
and the visibility squared $W(\vec\Theta,\lambda,B)$ are obtained 
using Eqs.~({\ref{Fdef}})--(\ref{Vaxisym}).

The relative figure of merit $R(\vec\Theta_m$,$\mathcal T$,$\mathcal 
N$,$B_j)$ and the robustness ratio $S(\vec\Theta_m$,$\mathcal 
T$,$\mathcal F$,$B_j)$ are calculated in two stages, implemented as 
separate programs.

First, for the given grids of model parameters
$\{\vec\Theta_m:m=1,\dots,N_{\mathrm m}\}$  and the projected 
baseline lengths
$\{B_j:j=1,\dots,N_B\}$, we calculate and store the matrices 
$\left(\tens C^{\mathrm S}\right) ^{-1}$ 
and 
$\left(\tens C^{\mathrm I}\right) ^{-1}$,
defined in Eqs.~(\ref{cspec})
and (\ref{colbi}). The defining parameters of the grids are
entered as input data, the grid of models being constructed
as the direct product of uniform grids for individual parameters.

The derivatives entering definitions of the matrices are approximated 
by finite differences.  All the model parameters that are not constant 
on the grid are treated as adjustable, that is either target or 
nuisance.  The distinction between those two types of parameters is 
irrelevant at this stage.

At the second stage, for given partitions of set $\mathcal T\cup\mathcal N$
of all adjustable parameters on the subsets {$\mathcal T$} and {$\mathcal N$},
we compute the values of  
$R(\vec\Theta_m$,$\mathcal T$,$\mathcal N$,$B_j)$ and
$S(\vec\Theta_m$,$\mathcal T$,$\mathcal F$,$B_j)$ 
(see Eqs.~(\ref{Rdef}) and  (\ref{Sdef}))
on the grid. In this way, the dependence of the results on
{$\mathcal T$} and {$\mathcal N$} can be studied without repeating 
the time consuming physical modeling of the envelope.
\section{Results \label{results}}
In total, the values $R(\vec\Theta,\mathcal T,\mathcal N,B)$ and 
$S(\vec\Theta,\mathcal T,\mathcal F,B)$ were calculated as a function 
of projected baseline length for various sets of target and
nuisance parameters at  
more than $10^5$ points of the parametric space.  Table \ref{grids} 
presents the principal characteristics of the model grids.  The ranges 
of variations of model parameters on the grids were chosen so as to 
include the values of parameters from earlier works on P~Cyg
cited in Sect.~\ref{stpars}.
\begin{table*}[tb]
\caption[]
 {Grids of models: the number of points in the
  parametric space $N_{\mathrm m}$;
  dimension, that is the number of varied parameters, of the grid
  $D$;
  the values of parameters that were fixed on the grid; and the
  ranges of variation for variable parameters. \label{grids}
 }
\begin{flushleft}
\begin{tabular}{lrllllllrrrrrrrr} 
\noalign{\smallskip}
\hline
\noalign{\smallskip}
Grid   &$N_{\mathrm m}$&$D$&$\alpha$ &$\dot M$&$v_{\infty}$&$v_{\mathrm c}$
       &$T_1$&$R_2$&$\Delta_2T$&$R_3$&$\Delta_3T$&$R_4$&$\Delta_4T$&$R_5$
       &$\Delta_5T$\\
no.    &         &       &         &$M_\odot\mbox{yr}^{-1}$
       &\mbox{km s$^{-1}$}&\mbox{km s$^{-1}$}&K &    &K  &    &K  
       &    &K  &    &K   \\
\hline\noalign{\smallskip}
1              &10\,000 &3&3.5 &1.0\,10$^{-5}$ &200 &15.0 &11\,000 \\
              &&
                           &4.5 &2.0\,10$^{-5}$ &    &55.0 &15\,000 \\
\noalign{\medskip}
2              &10\,000&3 &3.5 &1.0\,10$^{-5}$ &200 &15.0 &11\,000 
                &2.0 &1300 &3.4 &2100 &5.0 &200 &10.0 &3000\\
              &&
                           &4.5 &2.0\,10$^{-5}$ &    &55.0 &15\,000 \\
\noalign{\medskip}
3              &32\,805&4&3.5 &1.0\,10$^{-5}$ &150 &15.0 &11\,000 \\
              &&
                          &4.5 &2.0\,10$^{-5}$ &250 &55.0 &15\,000 \\
\noalign{\medskip}
4              &15\,625 &6&3.5 &1.0\,10$^{-5}$ &200 &15.0&11\,000
                                                           &2.0 &0 &5.0 &0 \\
              &&
               &4.5 &2.0\,10$^{-5}$ &    &55.0 &15\,000 &   &1300 &   &2300 \\
\noalign{\medskip}
5              &46\,656 &6&3.25 &1.0\,10$^{-5}$ &200 &15.0&11\,000
                                                           &2.0 &0 &3.4 &0 \\
              &&
               &4.25 &2.0\,10$^{-5}$ &    &55.0 &15\,000 &   &1300 &   &2100 \\
\noalign{\medskip}
6              &6561 &8&3.5 &1.0\,10$^{-5}$ &200 &15.0 &11\,000 
                &2.0 &0 &3.4 &0 &5.0 &0 &10.0 &0\\
              &&
                           &4.5 &2.0\,10$^{-5}$ &    &55.0 &15\,000
                                   &    &1300 & &2100 & &200 & &3000 \\     
\noalign{\medskip}
7              &65536 &8&3.5 &1.0\,10$^{-5}$ &200 &15.0 &11\,000 
                &2.0 &0    &3.0 &0    &4.0 &0   &5.0 &0\\
              &&
                      &4.5 &2.0\,10$^{-5}$ &    &55.0 &15\,000
                &    &1300 &    &2100 &    &200 &     &3000 \\     
\noalign{\medskip}
8              &6561 &8&4.0 &1.5\,10$^{-5}$ &180 &20.0 &14\,000 
                &2.0 &0 &3.0 &0 &4.0 &0 &5.0 &0 \\
              &&
                        &4.5 &2.3\,10$^{-5}$ &220 &     &16\,000
                &    &1000 & &1000 & &1000& &1000 \\
\noalign{\medskip}
9              &19\,683 &9&3.5 &1.5\,10$^{-5}$ &220 &15.0 &16\,000 
                &2.0 &0 &3.0 &0 &4.0 &0 &5.0 &0 \\
              &&
                       &4.5 &2.3\,10$^{-5}$ &300 &25.0 &18\,000
                &    &1300 & &2100 & &200 & &3000 \\     
\noalign{\medskip}
\hline
\end  {tabular}
\end  {flushleft}
\end  {table*}

The main result of the present study is that the increase in accuracy 
provided by OLBI strongly depends on the number of free parameters in 
the model: it varies from almost negligible for models of low 
dimension to very significant for models of high dimension.  Since 
in our case the boundary between ``low'' and ``high'' is nearly 
coincident with division onto models with fixed and adjustable thermal 
structure, we discuss the results pertaining to those two classes of 
models separately.
\subsection{Models with fixed thermal structure}
Among our grids, the grid~1 most closely corresponds to the models 
often used in interpreting H$\alpha$ observations of P~Cyg stars, 
where {$v_{\infty}$} is excluded from the set of adjustable parameters 
of the model (see e.g. Scuderi et al.  \cite{scuderi}) and its value 
is taken from the analysis of absorption lines of metals in the 
ultraviolet (Casatella et al.  \cite{vinf79}; Lamers et al 
\cite{vinf85}).

\begin{figure}
\resizebox{\hsize}{!}{\includegraphics{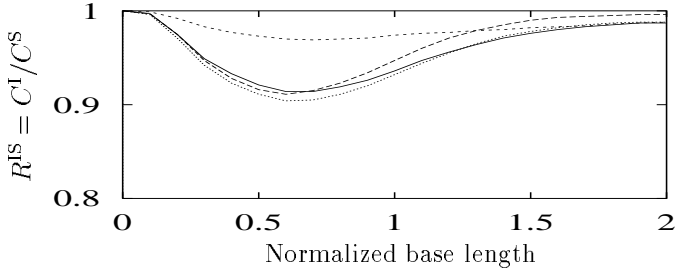}}
\caption[]
{Error ratio for isothermal model with 
 $\alpha=3.94$, $\dot M=1.44\,10^{-5}\,M_\odot\mbox{yr}^{-1}$,
 $v_{\mathrm c}=32.8{\mbox{ km s$^{-1}$}}$, $T_1=12778\,{\mathrm K}$, 
 $v_\infty=200{\mbox{ km s$^{-1}$}}$, and the set of adjustable parameters 
 $\mathcal T\cup\mathcal N=\{\alpha$,$\dot M$,$v_{\mathrm c}$,$T_1\}$.
 Target set 
 $\mathcal T=\{\alpha\}$ (solid line), $\{\dot M\}$ (dashed line), 
 $\{T_1\}$ (short-dashed line), and $\{v_{\mathrm c}\}$ (dotted line).
}
\label{isothtpc}
\end{figure}

The typical results for an individual point in the parametric space 
are illustrated by Fig.~\ref{isothtpc}, where the random error ratio
$R^{\mathrm{IS}}= C^{\mathrm I}(\vec\Theta,\mathcal T,\mathcal N,B)
                /C^{\mathrm S}(\vec\Theta,\mathcal T,\mathcal N,B)
$
is plotted as a function of the normalized baseline length $B/B_0$ for 
various {$\mathcal T$} and {$\mathcal N$}.  Here $B$ is the projected 
baseline length, $B_0=\lambda/2\pi\delta$, and $\delta$ is the angular 
radius of the central star.  For the adopted values of 
$R_*=76\,R_\odot$, $d=1800\,\mathrm{pc}$, and $\lambda=6265$\,\AA{}
one gets $B_0=110\,{\mathrm m}$.

The principal features of dependence of $R^\mathrm{IS}$ on $B$ are
independent on details of physical model used and can
be easily explained qualitatively. As shown in
Sect. \ref{RelFigMer}, inequality $R^\mathrm{IS}<1$ holds for all
$0<B<\infty$. When the projected baseline length is very small,
the interferometry does not provide any additional information as
compared with the spectroscopy, for the object gets
unresolved. Consequently, $R^\mathrm{IS} \to 1$ for $B \to 0$. In
the opposite case of large $B$ the fringe contrast is close to 0
for all physically realistic intensity distributions and its
dependence on model parameters is impossible to measure because of
observational errors. That is in that case the interferometry
again provides no additional information and 
$R^\mathrm{IS} \to 1$ for $B \to \infty$. 

It follows from the
above stated general properties of $R^\mathrm{IS}$ that for any
$\mathcal T$ and $\mathcal N$ it has a (possibly non-unique)
minimum at a certain finite value of $B$. The location
of the minimum indicates the projected baseline length at which
interferometric observations are most informative, and the
corresponding value of $R^\mathrm{IS}$ characterize to what degree
the interferometry can reduce the errors in model parameters
determination as compared to spectroscopy.

Let us note that it would be insecure to compare the 
merits of the methods using values of $R^{\mathrm{IS}}$ obtained for 
an isolated point or even for few arbitrarily chosen points of
parametric space. Since the model parameters are not known a
priori, it is necessary to use integral characteristics
describing the behavior of $R^\mathrm{IS}$ on the whole grid. We
will use the following three values calculated as a function
of $B_j/B_0$ for various {$\mathcal T$} and {$\mathcal N$}:
\begin{eqnarray}
R_{\min}(\mathcal T,\mathcal N,B_j)&=&
   \min_{1\le m\le N_{\mathrm m}}
        R^{\mathrm{IS}}(\vec\Theta_m,\mathcal T,\mathcal N,B_j)
                                              \label{rmindef}\\
R_{\max}(\mathcal T,\mathcal N,B_j)&=&
   \max_{1\le m\le N_{\mathrm m}}
        R^{\mathrm{IS}}(\vec\Theta_m,\mathcal T,\mathcal N,B_j)
                                               \label{rmaxdef}
\end{eqnarray}
\begin{equation}
P_\mathrm{opt}(\mathcal T,\mathcal N,B_j)=N_\mathrm{opt}/N_{\mathrm m},
                                               \label{poptdef}
\end{equation}
where {$N_\mathrm{opt}$} is the number of points on the grid for which 
function $R^{\mathrm{IS}}(\Theta,\mathcal T,\mathcal N,B)$ reaches its 
minimum at $B=B_j$.  

The function defined in Eq.~(\ref{poptdef})
is immediately related to the problem of optimal choice of the baseline 
length: for given {$\mathcal T$} and {$\mathcal N$}, the higher the 
value $P_\mathrm{opt}(\mathcal T,\mathcal N,B)$, the higher the 
probability that the interferometric observations at the baseline length 
$B$ would yield most accurate model parameters.
 
\begin{figure}
\resizebox{\hsize}{!}{\includegraphics{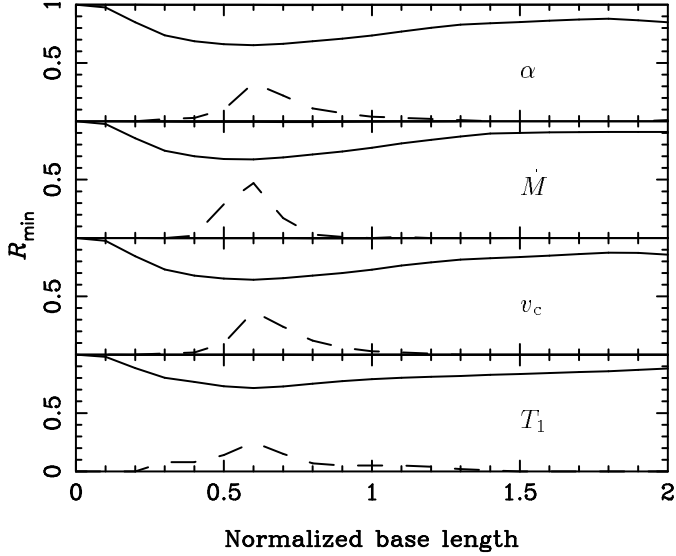}}
\caption[]
{
 The lower limit of random error ratio $R_{\min}$ (solid line) and
 distribution of optimal baseline lengths $P_\mathrm{opt}$ (dashed lines,
 plotted in an arbitrary scale) for grid~1.
 The set of adjustable parameters 
 $\mathcal T\cup\mathcal N=\{\alpha,\dot M$,$v_{\mathrm c},T_1\}$,
 target parameters are indicated on the plots.
}
\label{isothrdt}
\end{figure}

The dependence of {$R_{\min}$} and {$P_\mathrm{opt}$} on the
normalized baseline length  
and set of target parameters for grid~1 is shown in 
Fig.~\ref{isothrdt}.  For this grid, the value of {$R_{\max}$} is 
close to unity for all $\mathcal T$ and $\mathcal N$ and is not shown 
in the plot.
It can be easily seen that for this grid of models OLBI 
provides only relatively small reduction of the random error in parameter 
determination as compared with spectroscopy.  This conclusion remains 
also valid if we add the terminal velocity {$v_{\infty}$} to the set 
of adjustable parameters (grid 3).

As $R^{\mathrm{IS}}(\vec\Theta,\mathcal T,\mathcal N,B)\approx1$ for 
all $\mathcal T$ and $\mathcal N$, the same approximate equality is 
evidently valid for the robustness ratio $S(\vec\Theta,\mathcal 
T,\mathcal F,B) =R(\vec\Theta,\mathcal T,\mathcal 
F,B)/R(\vec\Theta,\mathcal T,\emptyset,B) $, which indicate (see 
Sect.~\ref{systerrs}) that, for models with a priori fixed thermal 
structure, interferometric data will neither reduce significantly the 
level of systematic errors.
\subsection{Models with adjustable thermal structure}
The number of possible target sets is an exponentially increasing 
function of the dimension {$D$} of the model parameter space, so that 
when the values $\Delta_iT$ are added to the set of unknowns, it 
becomes impossible to make an exhaustive presentation of even the 
integral results, i.e. the functions $R_{\min}(\mathcal T,\mathcal 
N,B)$,\\
$R_{\max}(\mathcal T,\mathcal N,B)$, and $P_\mathrm{opt}(\mathcal T,\mathcal N,B)$.

\begin{figure}
\resizebox{\hsize}{!}{\includegraphics{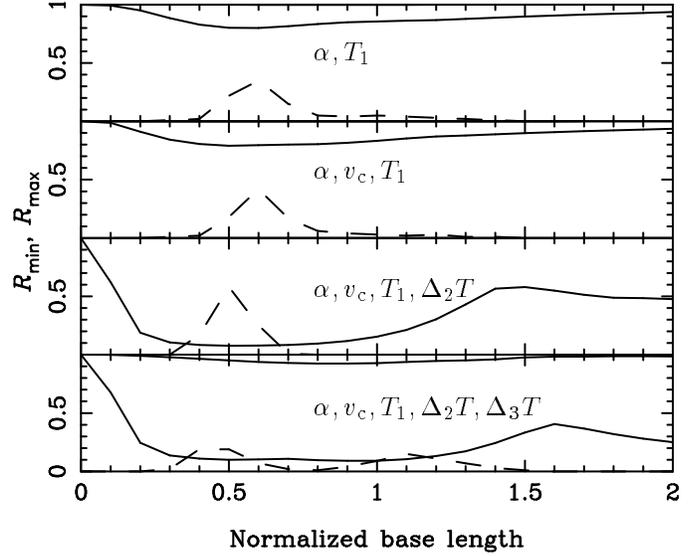}}
\caption[]
{
 The lower and upper limits of random error ratio (solid lines,
 $R_{\max}$ differs noticeably from unity only in the bottom plot) and
 distribution of optimal baseline lengths $P_\mathrm{opt}$ (dashed lines,
 plotted in an arbitrary scale) for grid~5.
 Target set $\mathcal T=\{\dot M\}$, nuisance parameters are indicated on the plots.
}
\label{3lmerr}
\end{figure}

In Fig.\ref{3lmerr}, we show the dependence of $R_{\min}$,
$R_{\max}$, and $P_\mathrm{opt}$ on 
the number of nuisance parameters for one of our simplest grids of 
models with adjustable thermal structure, the target set 
consisting of one parameter, the mass loss rate $\dot M$.  From many points 
of view, 
it is the most important parameter of the stellar wind, characterizing 
both its effect on the evolution of the star and the influence of the 
outflowing matter on the surrounding interstellar medium.

As it can be seen, 
the lower limit of the random error ratio $R_{\min}$ rapidly decreases as 
the number of adjustable temperature parameters increases, so that 
when the thermal structure of the envelope becomes a nuisance parameter, 
interferometry can reduce the random error in $\dot M$ by as much as an 
order of magnitude.

Note, that the decrease in $R_{\min}$ is caused by the fact that 
parameters $\Delta_iT$ are adjustable, rather then by deviation from 
isothermicity.  This can be seen from the upper two plots in 
Fig.~\ref{3lmerr}, where $R_{\min}>0.8$ for a grid containing high 
proportion of models with noticeable temperature gradient, and is also 
supported by the results obtained for grid~2, which differs from 
grid~1 only in that its thermal structure approximates that of Drew's 
model B, which corresponds to $\alpha=4$, 
$v_{\mathrm c}=15\,\mathrm{km/s}$, $v_\infty=300\,\mathrm{km/s}$,
and $\dot M=1.5\,10^5\,\,M_\odot\mbox{yr}^{-1}$ in our notations,
instead of being isothermal.

\begin{figure}
\resizebox{\hsize}{!}{\includegraphics{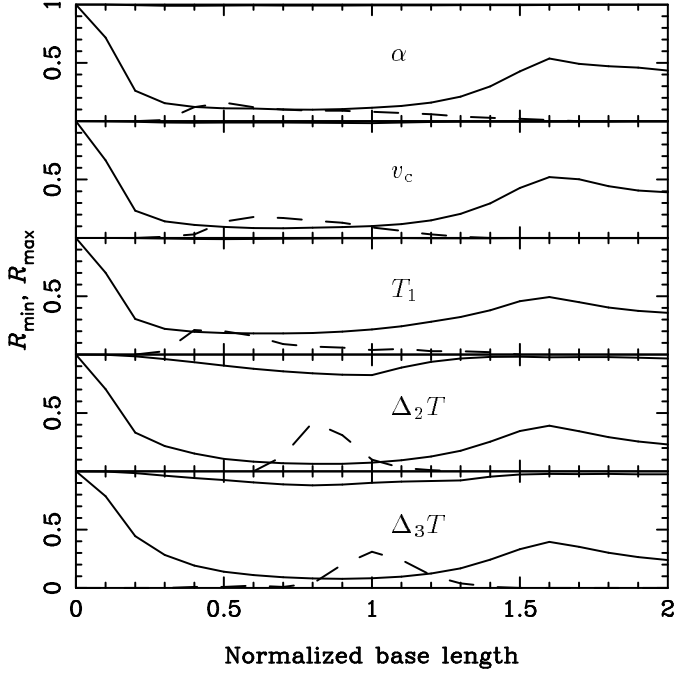}}
\caption[]
{
 The lower and upper limits of random error ratio (solid lines) and
 distribution of optimal baseline lengths $P_\mathrm{opt}$
 (dashed lines) for grid~5,
 $\mathcal T\cup\mathcal N$=$\{\alpha$, $\dot M$, $v_{\mathrm c}$,
 $T_1$, $ \Delta_2T$, $\Delta_3T\}$,
 and target sets indicated on the plots.
}
\label{3l1+5}
\end{figure}

For other single-parameter target sets the dependencies of $R_{\min}$ 
and $R_{\max}$ on $\mathcal N$ are qualitatively the same.  
Fig.~\ref{3l1+5} displays the results for the most important case when 
all the parameters of the model except for $v_{\infty}$ are 
adjustable.  As one can see, again, in favorable circumstances 
interferometry can increase accuracy by an order of magnitude.
In contrast to $R_{\min}$, the value $R_{\max}$ is close to unity for 
all projected baseline lengths and combinations of sets $\mathcal
T$ and $\mathcal N$ that we have studied. This signifies that for
any projected baselength there exists a combination of model
parameters for which interferometry practically do not provide
reduction in error. 

\begin{figure}
\resizebox{\hsize}{!}{\includegraphics{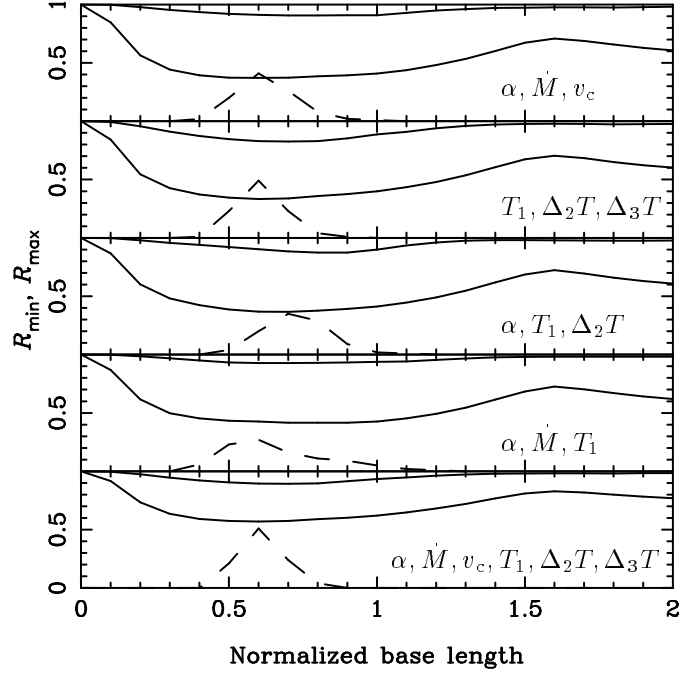}}
\caption[]
{
 The lower and upper limits of random error ratio (solid lines) and
 distribution of optimal baseline lengths $P_\mathrm{opt}$
 (dashed lines) for grid~5,
 $\mathcal T\cup\mathcal N$=$\{\alpha$, $\dot M$, $v_{\mathrm c}$,
 $T_1$, $ \Delta_2T$, $\Delta_3T\}$,
 and target sets comprising several parameters (indicated on the plots).
}
\label{3lmultyt}
\end{figure}

It is interesting to notice that for target sets containing several 
parameters (see
Fig.~\ref{3lmultyt}), the influence of individual parameters on 
$R_{\min}$ and $R_{\max}$ to some extent averages out: $R_{\min}$ 
is systematically higher, and $R_{\max}$ is systematically lower than 
the corresponding values for individual parameters.  For this reason, 
when more than one model parameter is to be determined,
interferometry at a single projected baselength is unlikely to
yield overall gain in accuracy in excess of factor of two even at
optimal baselines. 

We studied further the effect of increasing the number of sublayers
$M_R$, which permits to investigate the influence of fine details of
the thermal   
structure of the envelope. This appears to be justified for 
Drew (\cite{drew}) showed that in certain cases this structure 
is quite complex, and gradient of $\log T(r)$ varies 
strongly with radial distance.  Consequently, a model aiming to closely
approximate the realistic envelope should be parameterized using 
a large $M_R$ (see Sect.~\ref{stpars})
and hence requires computations on grids of high dimension $D$.

Since the number of mesh points of the grid depends on {$D$} 
exponentially, for computations based on such complex models (grids 
5--9 from Table~\ref{grids}) the limited computer resources 
dictate sparser mesh points along each axis of the parameter space and 
a tuning of parameters controlling the computational process in such a way 
as to increase the speed of computation at the expense of precision.

\begin{figure}
\resizebox{\hsize}{!}{\includegraphics{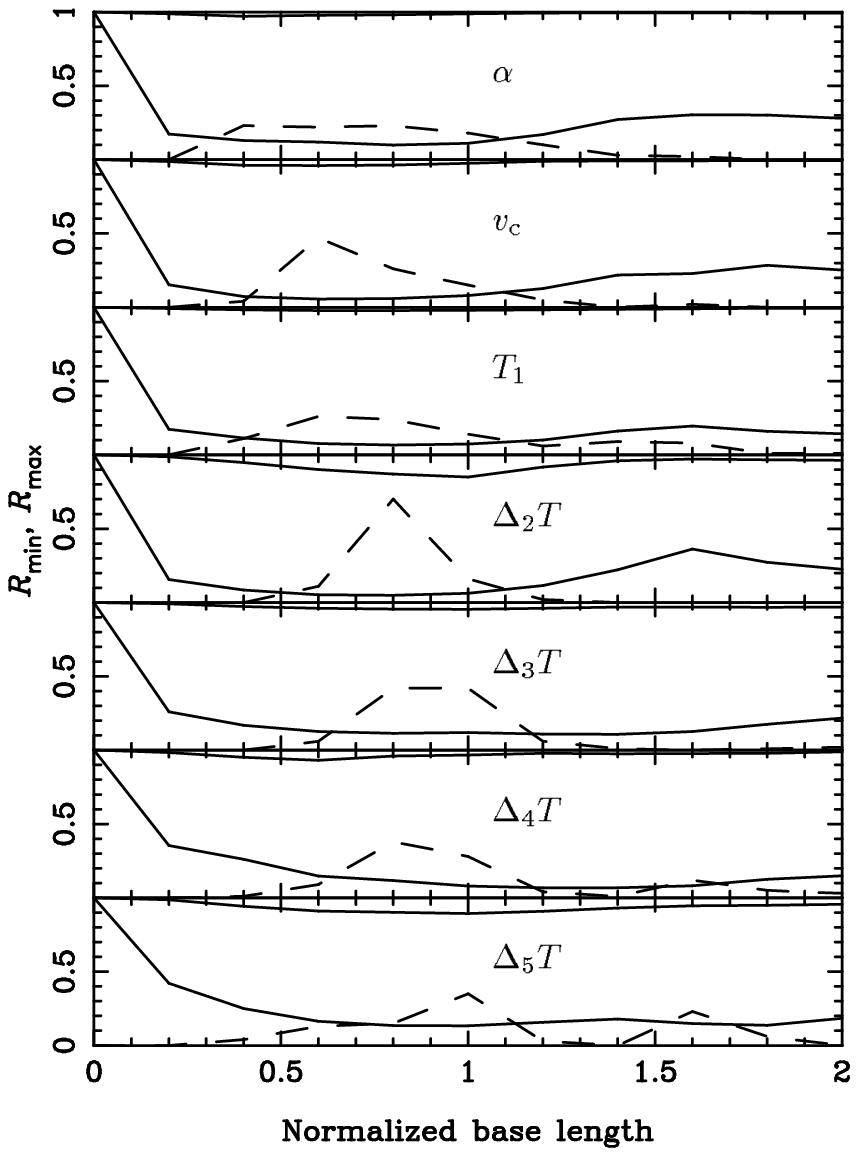}}
\caption[]
{
 The lower and upper limits of random error ratio (solid lines) and
 distribution of optimal baseline lengths $P_\mathrm{opt}$ 
 (dashed lines) for grid 7,
 $\mathcal T\cup\mathcal N$=$\{\alpha$, $\dot M$, $v_{\mathrm c}$,
 $T_1$, $ \Delta_2T$, $ \Delta_3T$, $ \Delta_4T$, $ \Delta_5T\}$,
 and target sets consisting of a single parameter (indicated on
 the plots).
}
\label{5l1+7}
\end{figure}

Although numerically less precise than computations for simpler
models, our results for  more realistic grids indicate
that the conclusions derived for low
$M_R$ can be safely extrapolated for higher $M_R$.
This point is illustrated in Fig.~\ref{5l1+7} showing the results
for single-parameter target sets for grid~7.
As it can be seen in Figs.~\ref{3lmerr}--\ref{5l1+7} and is confirmed 
by our other results not presented here, the distribution of optimal 
projected baseline lengths in the majority of cases has a maximum
in the range  
0.4--0.8 in dimensionless units.  In some cases this maximum shifts to 
higher baseline lengths, but this occurs only for a very special choice of 
$\mathcal T$ that correspond to measurements of wind temperature 
gradient at the distances of several stellar radii from the star with 
all other model parameters considered nuisance (see e.g. bottom plots 
in Figs.~\ref{3l1+5} and \ref{5l1+7}).  Thus, the aforementioned 
range, corresponding to 45--90~m in linear scale, can be recommended 
as the optimal choice for interferometric observations aimed at 
determination of global characteristics of the P~Cyg wind.

\begin{figure}
\resizebox{\hsize}{!}{\includegraphics{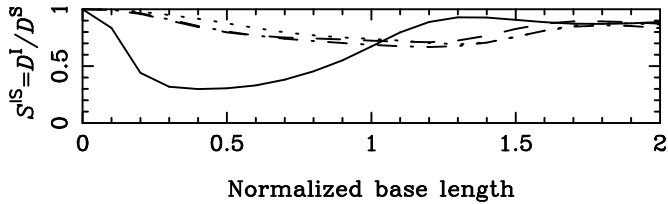}}
\caption[]
{ Robustness ratio for the model
  $\alpha=4.00$, $\dot M=1.5\,10^{-5}\,M_\odot/\mathrm{yr}$, 
  $v_{\mathrm c}=35.0{\mbox{ km s$^{-1}$}}$, $T_1=13000\,{\mathrm K}$, 
  $\Delta_2T=650\,{\mathrm K}$, 
  $\Delta_3T=1050\,{\mathrm K}$, $\Delta_4T=100\,{\mathrm K}$, 
  $\Delta_5T=1500\,{\mathrm K}$
  from grid~6 and $\mathcal T$=\{$\dot M$\}.
  Set of fixed parameters 
  $\mathcal F=\{\alpha$, $v_{\mathrm c}$, $ T_1$, $\Delta_2T\}$ (solid line), 
  $\{\alpha$, $v_{\mathrm c}$, $ T_1$, $ \Delta_2T$, $\Delta_3T\}$ 
   (dashed line),
  $\{\alpha$, $v_{\mathrm c}$, $ T_1$, $ \Delta_2T$, $ \Delta_3T$, 
     $ \Delta_4T\}$ (dash-dotted line), 
   and
  $\{\alpha$, $v_{\mathrm c}$, $ T_1$, $ \Delta_2T$, $ \Delta_3T$, 
   $ \Delta_4T$, $\Delta_5T\}$ (dotted line).
}
\label{rbst}
\end{figure}
The typical dependence of the robustness ratio \\
$S^{\mathrm{IS}}(\vec\Theta,\mathcal T,\mathcal N,B) = U^{\mathrm I}/U^{\mathrm S}$
on the baseline length and the set of fixed parameters for a model with 
adjustable thermal structure is shown in Fig.~\ref{rbst}.  At nearly 
all points of parametric space $S^{\mathrm{IS}}<1$, and $S^{\mathrm 
IS}$ is generally lower when the set of fixed parameters $\mathcal F$ 
includes the values $\Delta_iT$.  For a small fraction of 
models and for short projected baselines where interferometry can not significantly 
reduce random error, the value $S^{\mathrm{IS}}>1$ by a negligible 
amount.

Thus, our results indicate that if the baseline length and the model 
used for interpretation are chosen in such a way as to reduce random 
errors, OLBI appears also able to reduce the systematic errors.
\section{Conclusions}
Using the theory of model parameter determination, we developed a 
general method which permits a quantitative comparison of 
observational techniques and optimization of complex observations 
through the relative figure of merit, defined as 
a generalization of the ratio of random errors of model parameters.

The method was applied to compare the outcome of OLBI and classical 
spectroscopic observations, both used to determine the parameters of 
the outflow of the P~Cyg star.  The observable quantities were 
computed using an efficient radiative transfer code and realistic 
grids of the envelope model.  The OLBI and spectroscopic measurements 
errors were simulated assuming that the only source of errors is the 
counting statistics, i.e. we considered a nearly-ideal interferometer.  
Other main assumptions and simplifications of the present work were as 
follows: (1) the P~Cyg wind is stationary and spherically symmetric; 
(2) the only OLBI observable considered here is the visibility modulus; 
(3) only the hydrogen H$\alpha$ line is considered.

The main conclusions of the present work are as follows:
\begin{enumerate}
\item
A meaningful evaluation of the OLBI vs spectroscopy relative figure of 
merit requires exploration of large domain of the model parametric 
space (about $10^5$ points computed in the present work).
\item
If the P~Cyg outflow can be accurately described by a 
simple model of a stationary spherically symmetric isothermal 
envelope, then the use of interferometry does not substantially reduce the random error in the model 
parameter determination.
\item
If the P~Cyg outflow requires more complex models, e.g. with a priori unknown
dependence of the envelope temperature on radial distance, use of a
nearly-ideal interferometer can reduce errors in model parameters by an 
order of magnitude as compared to the spectrometer of the same 
collecting power and noise level.
\item
The study of the relative robustness, i.e. the stability of the 
obtained model parameter determination with respect to systematic 
biases in the model, indicates that when the OLBI and the physical 
model reduce random errors, they also tend to reduce the systematic 
error.
\item
The optimal projected baseline lengths for observation of P~Cyg with a nearly ideal
interferometer lie in the range 45--90 m.
\item
If several model parameters are to be determined simultaneously,
the OLBI observations at a  single projected baseline cannot provide a substantial error 
reduction.
\end{enumerate}
\section{Final remarks}
In this section we discuss to what extent our results are relevant for 
real observations, and how the methods developed here could be 
generalized to more complex cases.
\subsection{Additional sources of measurement errors}
Although general formalism presented in
Sect.~\ref{statth} remains valid for any source of errors
provided that the linearized least-square method is applicable,
Eqs.~(\ref{cspec}) and (\ref{colbi}) need to be modified if the
accuracy of measurements is not shot noise limited.

In the simplest case of uncorrelated errors in spectral channels,
this modification reduces to multiplication of the integrands in
Eqs.~(\ref{cspec}) and (\ref{colbi}) by corresponding ratios 
(shot noise error)/(total error). If this ratios are nearly
constant across the observed spectral range, then after
appropriate scaling our results can be directly applied to real
observations.

The assumption of uncorrelated errors may be invalid for 
interferometric observations,  the observables $\hat W_i$ being obtained 
as a result of much more complicated processing of raw observational 
data (cf. Mourard et al \cite{mour94}) then that considered in 
\ref{olbidiag}.  
This point must be further analysed, and in future works, it may 
be necessary to use directly Eq.~(\ref{covmateval}) instead of 
Eq.~(\ref{colbi}).

In any case, a generalization on arbitrary measurement errors could be 
easily incorporated into existing code with no considerable increase 
in the required computer resources.
\subsection{Observation in several spectral lines}
The theory developed in Sect.~\ref{statth} can be applied to the
case of multiple observed lines without any
modifications. Formally, it reduces to changing
integration domains in Eqs.~(\ref{cspec}) and (\ref{colbi}).

Rapid evaluation of source function, on the contrary, would
require a detailed study and development of efficient methods
specific to each spectral line involved, because different lines
can be emitted in different regions of the envelope and originate
from different physical processes.
\subsection{Multi-baseline OLBI of spherically symmetric objects}
The method used here can be generalized to visibility
measurements done simultaneously at more than one baseline: the only modification required is  
the replacement of RHSes of Eq. (\ref{colbi}) by 
the sum of corresponding expressions over given set of projected
baseline lengths, the directions of the baselines being
unimportant because of circular symmetry of intensity
distribution. 
This can be done by modifying only the code performing the second 
stage of calculations (see Sect.~\ref{compfinres}) and using the 
data prepared for the single-baseline case during the most time consuming 
first stage of computation.
\subsection{Asymmetric brightness distribution and
information on fringe phase}
Although time-averaged structure of P~Cyg wind can be considered
spherically symmetric, polarimetric (Taylor et al.
\cite{asymwind}) and interferometric (Vakili et
al. \cite{gi2tpcyg}) observations revealed existence of
time-dependent "clumpy" structures in the envelope, giving rise to 
deviation from spherical symmetry.

In this case,
measurements of the fringe phase can yield important information.  This 
contrasts with the spherically symmetric case, where the fringe phase is 
irrelevant to physical modeling of the object.  

On the other hand, a unique determination of model parameters using only spatially 
unresolved spectroscopy appears impossible for the complex physical models 
resulting in two-dimensional brightness distributions, so that  
even the formulation of the  comparison problem for spectroscopy and 
interferometry can be hardly done.  The choice of the optimal set of 
baseline lengths for interferometry still remains the important problem, 
and the method developed here can be used for such an optimization 
after certain modifications.

First, the fringe phases (or observable linear combinations thereof)
should be included into the set of observables defined in
Eq.~(\ref{Yolbi}), and Eq.~(\ref{colbi}) should be modified
accordingly. This would require comprehensive error analysis
depending on particular method used for extracting phase
information from measurements.

Second, errors in parameter determination for observations performed 
with interferometer configurations $B_1$ and $B_2$ (here $B_1$ and 
$B_2$ denote corresponding sets of simultaneously used projected baseline lengths) 
can be compared using the ratio \\
$C^{\mathrm I}(\vec\Theta,\mathcal T,\mathcal N,B_1)
/C^{\mathrm I}(\vec\Theta,\mathcal T,\mathcal N,B_2)
$
instead of $R^{\mathrm{IS}}(\vec\Theta,\mathcal T,\mathcal N,B)$.
\begin{acknowledgements}
The authors are grateful to A. Labeyrie, D. Mourard and F.Vakili for 
encouraging discussion at the beginning of this project.  We thank 
C.Bertout who provided the code for computation of the emergent 
intensities.  MSB is grateful to the staff of Observatoire de Grenoble 
for the hospitality during his stay in Grenoble.  This research made 
use of SIMBAD astronomical database, created and maintained by the 
CDS, Strasbourg, and NASA's Astrophysics Data System Abstract Service.  
Part of this work was supported by CNRS through the grants PICS 
${\mathrm n}^0$~194 ``France -- Russie, Astronomie \`a haute 
r\'esolution angulaire", GdR ``Milieux circumstellaires", and GdR and 
PN ``Haute r\'esolution angulaire en Astronomie".
\end{acknowledgements}
\appendix
\section{Computation of ${\tens D}{w_i}$ and ${\tens M}(n_i w_i)$ 
         \label{olbidiag}}
In this Appendix, we compute the non-trivial elements of the
covariance matrix of errors used in Sect.~\ref{interferometry}.
Since we consider here a single spectral channel, the index $i$
will be omitted everywhere. Thus, the values $N$, $\hat{N}$, $n$, $W$,
$\hat{W}$, and $w$ in this Appendix are identical with the values
denoted as $N_i$, $\hat{N_i}$, $n_i$, $W_i$,
$\hat{W_i}$, and $w_i$ respectively in Sect.~\ref{statth}.

In practice, the error in {$\hat W$} depends on too many details of 
experimental techniques, atmospheric conditions and data processing 
methods to be analyzed in general form (Mourard et al \cite{mour94}).  
The present analysis is restricted to a highly idealized situation.  
Namely, as well as in Sect.~\ref{spectroscopy}, the only source of 
measurement errors taken into account is the photon shot noise.  We 
assume that the static fringes are detected by counting the photons in 
nonintersecting channels that cover in total the interval of length 
$Z$ of optical path difference (OPD).
If z is the OPD, which is measured here in units of $\lambda/2\pi$, and 
$F(z)$ is the photon counting rate per unit $z$, then
\begin{equation}
 F(z)=N(1+V\cos(z+z_0))/Z\,,                      \label{Vdef}
\end{equation}
where  $z_0$ is unknown fringe phase shift and
$N$ is proportional to the total received flux.

If we define
\begin{equation}                                  \label{Fcos}
 C=2\int_0^ZF(z)\cos z\,{\mathrm d}z
\end{equation}
and
\begin{equation}                                  \label{Fsin}
 S=2\int_0^ZF(z)\sin z\,{\mathrm d}z\,,
\end{equation}
then assuming that $Z\gg 1$ and neglecting the  
terms that decrease as $1/Z$ for $Z\to\infty$
we obtain that 
\begin{equation}                                   \label{Nolbi}
N=\int_0^ZF(z)\,{\mathrm d}z
\end{equation}
and
\begin{equation}                                   \label{Wvalue}
V^2=(C^2+S^2)/N^2\,.
\end{equation}
An estimate for $W=V^2$ can be constructed by replacing the values in 
the RHS of Eq.~(\ref{Wvalue}) by their estimates that can be easily 
deduced from Eqs.~(\ref{Fcos})--(\ref{Nolbi}).  If the interval of 
OPDs covered by measurements is divided in {$M$} subintervals of 
length $\Delta_p z$ and $\Delta_p z\ll1$ for $p=1,\ldots,M$, then 
estimate {$\hat W$} of $W$ based upon the number of photons received 
in each of these subintervals is given by
\begin{displaymath}
   \hat W=(\hat C^2+\hat S^2)/\hat N^2\,,
\end{displaymath}
where 
$\hat N=\sum_{p=1}^{M}{\hat N}_p$,
${\hat N}_p$ is the number of photons received in the
$p$-th subinterval,
  $\hat C=2 \sum_{p=1}^{M}{\hat N}_p\cos z_p$, 
and 
  $\hat S=2 \sum_{p=1}^{M}{\hat N}_p\sin z_p$.

If we assume further that 
{$\hat W$} as a function of $n_p={\hat N}_p-N_p$
can be linearized in the vicinity of zero and take into account that 
  ${\tens M}(n_p)=0$
and
  $ {\tens M}(n_p n_q)=\delta_{pq}N_p$,
then after some algebra we obtain for the statistical
characteristics of $\hat W$ 
\begin{eqnarray*}
{\tens D}(w)&=& {\tens M}
                 \left( 
                  \left( 
                    \sum_{p=1}^{M}\frac {\partial W} {\partial  N_p} n_p 
                  \right)^2 
                 \right)
                = \frac{4W}{N}(2-W)\,,\\
{\tens M}(n w) &=&{\tens M}
                    \left(
                     \left(
                      \sum_{p=1}^{M}\frac {\partial W} {\partial  N_p} n_p 
                     \right)
                     \left( \sum_{q=1}^{M} n_q \right) 
                    \right)
                  =0\,,
\end{eqnarray*}
which proves the validity of Eqs.~(\ref{wwcovar}) and (\ref{nwcovar})
for $i=j$.
\section{Simple approximation for the escape probability  \label{epapprox}}   
In the Sobolev approximation, the escape probability for a line photon 
in the spherically symmetric accelerating outflow is given by (See 
e.g. Mihalas \cite{mihalas}, Sect.~14.2)
\begin{equation}                           \label{epdef}
\beta=\int_0^1\frac{1-\exp(-\chi_{\mathrm l}/Q(\mu))}
                   {\chi_{\mathrm l}/Q(\mu)}
              {\mathrm d}\mu\,,
\end{equation}
where
\begin{displaymath}
  Q(\mu)=\mu^2\partial{V}/\partial{r}+(1-\mu^2)(V/r)\,,
\end{displaymath}
\begin{displaymath}
\chi_{\mathrm l}=\frac{(\pi e^2/mc)f}{\Delta\nu_D}n\,,
\end{displaymath}
$V=v/v_\mathrm{th}$, $v$ and {$v_\mathrm{th}$} 
are the outflow and scattering atoms
thermal velocities respectively,
$\Delta\nu_D=\nu v_\mathrm{th}/c$, $\nu$ is the line frequency,
$f$ is the oscillator strength, and $n$ is the number density
of atoms on the lower level of the transition.

Here and in what follows we consider the fixed point of the
envelope at the distance $r$ from the center and do not show
explicitly the dependence of $\beta$, $V$, $n$, and other physical
parameters of the envelope on $r$. Also, in expression for {$\chi_{\mathrm l}$}
we neglected the stimulated emission, which do not introduce a
noticeable error when the formation of H$\alpha$ in the P~Cyg
envelope is considered.

Our purpose is to find an approximation {$\tilde\beta$} to {$\beta$} 
that permits fast evaluation.  In order that in solving the equations 
of statistical equilibrium the approximation could be used instead of 
exact definition, the former should retain principal mathematical 
properties of the latter reflecting underlying physics.  Namely, we 
require that $\tilde\beta(0)=1$ and $\tilde\beta(n)$ should 
monotonically decrease to zero as $n\to\infty$.

The simplest function that obey the above stated  requirements is
\begin{equation}                \label{epappdef}
\tilde\beta=\frac{1}{1+n/n_{\mathrm{as}}}\,,
\end{equation}
where {$n_{\mathrm{as}}$} is a parameter.
To choose the optimal value of $n_{\mathrm{as}}$, we consider the behavior 
of $\beta(n)$ for large $n$. When $n\to\infty$, the 
numerator of the integrand in
Eq.~(\ref{epdef}) can be replaced by 1 and asymptotic form of
{$\beta$} can be easily computed. 

Comparison of that asymptotic form
with Eq.~(\ref{epappdef}) shows that if we set
\begin{equation}                                        \label{nas}
n_{\mathrm{as}}= 8\pi\frac{g_{\mathrm l}}{g_{\mathrm u}}\frac{1}{\lambda^3}
      A_\mathrm{ul}(\frac{1}{3}\frac{dv}{dr}+\frac{2}{3}\frac{v}{r})\,,
\end{equation}
then $\tilde\beta/\beta=1$ up to the terms of order $1/n$ when 
$n\to\infty$.
Here $g_{\mathrm l}$ and $g_{\mathrm u}$ are statistical weights of the
lower and upper levels of the transition respectively,
$\lambda$ is the line wavelength, and $A_\mathrm{ul}$ is the
spontaneous emission coefficient.

To study the precision of approximation (\ref{epappdef}), we
performed numerical computations for a wide range of values
$(dv/dr)/(v/r)$ and $n/n_{\mathrm{as}}$  (it can be easily shown that the
error is uniquely determined by the values of these two
parameters).
Our results show that the 
relative error reaches its maximum of 0.23 for
 $\frac{{\mathrm d}v(r)}{{\mathrm d}r}=\frac{v(r)}{r}$,
that is when $v(r)\propto r$, and
 $n_2/n_{\mathrm{as}}\approx1.8$.

The approximation found here is applicable if the escape
probability is given by Eq. (\ref{epdef}), that is if the
frequency change in scattering is described by complete frequency
redistribution. Of course, it is useful only if the usual
conditions of applicability of the Sobolev approximation are
satisfied. 
The expressions (\ref{epappdef}) and (\ref{nas}) can be generalized in
two ways.  First, the error can be reduced if rational
approximations of higher order are used instead of
Eq.~(\ref{epappdef}).  Second, a similar approximation can be
obtained for a general three-dimensional flow as  
it will be showed in a future work.

\end{document}